\documentclass[twocolumn,showpacs,preprintnumbers,amsmath,amssymb,10pt]{revtex4}
\usepackage{psfig}
\usepackage{graphicx}% Include figure files
\usepackage{dcolumn}% Align table columns on decimal point
\usepackage{bm}% bold math
\newcommand{\s}{\ensuremath{\psi(t,r)}}
\newcommand{\n}{\ensuremath{\nu(t,r)}}
\newcommand{\T}{\ensuremath{\theta}}

\newcommand{\pt}{\ensuremath{p_\theta}}

\begin{document}

\preprint{}
\title{Black hole formation in perfect fluid collapse}
\author{Rituparno Goswami}
\email{goswami@tifr.res.in}
\author{Pankaj S Joshi}
\email{psj@tifr.res.in}
\affiliation{ Department of Astronomy and Astrophysics\\ Tata
Institute of Fundamental Research\\ Homi Bhabha Road,
Mumbai 400 005, India}

\begin{abstract} We construct here a special class of perfect fluid 
collapse models which generalizes the homogeneous dust collapse solution
in order to include non-zero pressures and inhomogeneities into
evolution. It is shown that a black hole is necessarily generated as 
end product of continued gravitational collapse, rather than a naked 
singularity. We examine the nature of the central singularity forming 
as a result of endless collapse and it is shown that no non-spacelike 
trajectories can escape from the central singularity. Our results provide 
some insights into how the dynamical collapse works, and into the possible 
formulations of the cosmic censorship hypothesis, which is as yet a 
major unsolved problem in black hole physics.

\end{abstract}
\maketitle

\noindent Black hole physics has attracted considerable attention
in recent years and has witnessed rapid theoretical developments as 
well as numerous astrophysical applications. It is to be noted, however, 
that while few exact static or stationary models of black holes such as 
the Schwarzschild, Reissner-Nordstrom, and Kerr-Newman spacetimes are 
well-studied, the actual {\it formation} of black holes within 
the framework of a {\it dynamical} gravitational collapse process is 
not really an arena where much is known.

In a realistic physical scenario, (stellar mass) black holes will 
be typically born when a massive star exhausts its nuclear fuel, and
then collapses endlessly under the influence of its own gravitational 
field. Towards modeling such a physical process, a well-known model 
that has served as the basic paradigm in black hole physics is that of 
the Oppenheimer-Snyder spherically symmetric collapse solution
\cite {os}, 
where a dust cloud undergoes a continued  
collapse to form a black hole. Here the collapse initiates from a regular
initial data, when there is no trapping of light (i.e. light rays
from the star can escape to
faraway observers). Subsequently, as the collapse advances 
the process of formation of an
event horizon and closed trapped surfaces takes place, thus leading to
the formation of black hole and the eventual spacetime singularity. 
The trapped surfaces and the event horizon form here well in advance
to the epoch of the formation of the spacetime singularity, which
is hence necessarily hidden within the black hole.

Even though this collapsing Friedmann model already tells 
us a homogeneous dust collapse will always end in a black hole
rather than a naked singularity, it should be noted that this 
scenario in fact has several limitations. For example,
in this case the cloud has no 
pressures included, whereas any physically realistic collapse must 
include pressures. 
Another restrictive assumption here is that the density profile 
is assumed to be strictly homogeneous in space, at all times through
out the evolution of the cloud. For any isolated object such as a star, 
one may rather like to study a physically realistic density 
distribution which would be typically higher at the center, 
and decreasing as we move away from the center.

It is thus essential to study and analyze more general collapse
situations in order to understand the black hole formation in more 
realistic
stellar collapse scenarios. This is also essential in order to make any 
possible progress towards the {\it cosmic censorship hypothesis} 
\cite{pen}, 
which broadly 
states that any physically realistic gravitational collapse must 
result into the development of  
a black hole. Such a conjecture has been absolutely 
fundamental to the theory of black holes, and has played a major 
role in astrophysical applications of the black hole physics. 
This, however, remains a major unresolved open problem in general 
relativity and black hole physics today.

From such a perspective, we study here a specific class of 
collapse models where matter obeys the perfect fluid equation of 
state, and construct models where the collapse always necessarily ends 
in the formation of a black hole. The models we study here are 
somewhat special in that the mass function is assumed to be 
separable in the variables which are the physical radius of the cloud, 
and the time
coordinate. However, this is a class which generalizes 
the Oppenheimer-Snyder dust collapse models in two important
respects, namely inhomogeneities of density distribution are
included and also non-zero pressures have been incorporated now. 
As the collapse always ends here in a black hole formation 
as we show, it is hoped that dynamical considerations such as these 
will provide some useful 
insights into physically realistic collapse and the actual 
process of black hole formation. 
It is not unlikely that it is only such dynamical considerations 
which would prove essential to resolve the issue of cosmic censorship.
The model here may be
of interest as it includes pressures which may be
important in the later stages of collapse, and because the equation
of state is that of a perfect fluid, which is physically a
well-studied equation of state widely used in 
various astrophysical considerations.

The fluid content of the cloud is in the form of 
a perfect fluid
with an equation of state of the form $p=k\rho$, {\it i.e.} at 
all epochs the radial and tangential pressures are equal and isotropic, 
and are proportional to the density function of the cloud. 
Though the case of a general inhomogeneous {\it dust} collapse 
with $k=0$ can be completely solved 
\cite{LTB}, 
there are still a number of open questions regarding the 
end state of a general perfect fluid collapse.
Our purpose here is to examine a class of solutions of 
the Einstein equations 
for a spherically symmetric perfect fluid to understand  
explicitly how an 
inhomogeneous density profile should behave in the later stages of 
collapse and near the singularity, so that the final state
of the collapse would always be a black hole necessarily.

The spacetime geometry within the spherically symmetric collapsing 
cloud can be described by the metric in the comoving   
coordinates $(t,r,\theta,\phi)$ as given by,
\begin{equation}
ds^2=-e^{2\n}dt^2+e^{2\s}dr^2+R^2(t,r)d\Omega^2
\label{eq:metric}
\end{equation}
where $d\Omega^2$ is the line element on a two-sphere. 
The energy-momentum tensor for any matter fields of {\it type I} 
\cite{haw}
(this is a broad class which includes 
most of the physically reasonable matter fields, including 
dust, perfect fluids, massless scalar fields and such
others) is then given in a diagonal form,  
\begin{equation}
T^t_t=-\rho(t,r);\; T^r_r=p_r(t,r);\; T^\T_\T=T^\phi_\phi=p_\T(t,r)
\label{eq:emtensor}
\end{equation}
The quantities $\rho$, $p_r$ and $p_\T$ are the energy density, radial 
and tangential pressures respectively of the cloud. We take the matter 
fields 
to satisfy the {\it weak energy condition}, i.e. the energy density 
measured by any local observer is non-negative. Then for any 
timelike vector $V^i$, we must have,
\begin{equation}
T_{ik}V^iV^k\ge0
\end{equation}
which amounts to,
\begin{equation}
\rho\ge0;\; \rho+p_r\ge0;\; \rho+p_\T\ge0
\end{equation}

Now for the metric (\ref{eq:metric}) the Einstein equations 
take the form, in the units $(8\pi G=c=1)$
\begin{equation}
\rho=\frac{F'}{R^2R'};\;\;\;\;p_r=-\frac{\dot{F}}{R^2\dot{R}}
\label{eq:ein1}
\end{equation}
\begin{equation}
\nu'=\frac{2(\pt-p_r)}{\rho+p_r}\frac{R'}{R}-\frac{p_r'}{\rho+p_r}
\label{eq:ein2}
\end{equation}
\begin{equation}
-2\dot{R}'+R'\frac{\dot{G}}{G}+\dot{R}\frac{H'}{H}=0
\label{eq:ein3}
\end{equation}
\begin{equation}
G-H=1-\frac{F}{R}
\label{eq:ein4}
\end{equation}
where,
\begin{eqnarray}
G(t,r)=e^{-2\psi}(R')^2; && H(t,r)=e^{-2\nu}(\dot{R})^2
\label{eq:ein5}
\end{eqnarray}

The arbitrary function $F=F(t,r)$ here has an 
interpretation of the mass function for the cloud, and it
gives the total mass in a 
shell of comoving radius 
$r$ on any spacelike slice $t=const$. We have $F\ge0$ from the energy 
conditions. In order to 
preserve the regularity at the initial epoch, we have  
$F(t_i,0)=0$, that is the mass function should vanish at the center 
of the cloud. Since we are considering collapse, we have $\dot R<0$,
i.e. the physical radius $R$ of the cloud keeps decreasing and
ultimately reaches $R=0$.
As seen from equation (6), there is a density singularity
in the spacetime at $R=0$, and at $R'=0$. However, the later ones
are due to shell-crossings and these weak singularities 
can be possibly removed from the spacetime 
\cite{cjsc}, 
so we shall consider here only the shell-focusing 
singularity at $R=0$, which is the genuine physical singularity where all
matter shells collapse to a zero physical radius.

Now let us incorporate the perfect fluid form 
of matter, where the
radial and tangential pressures are equal, and take the 
equation of state for
the collapsing matter to be 
\begin{equation}
p_r(t,r)=p_\T(t,r)=k\rho(t,r)
\label{eq:pf}
\end{equation}
where $k<1$ is a constant. Then the equations (\ref{eq:ein1}) 
and (\ref{eq:ein2}) become,
\begin{equation}
\rho=\frac{F'}{R^2R'}=-\frac{1}{k}\frac{\dot{F}}{R^2\dot{R}}
\label{eq:ein6}
\end{equation}
\begin{equation}
\nu'=-\frac{k}{k+1}\left[\ln(\rho)\right]'
\label{eq:ein7}
\end{equation}
Thus we see that there are five dynamical variables, 
namely  $\rho$, $\psi$, $\nu$, $R$, and $F$, and there are five 
total field equations. Also,
using the scaling independence we can write $R(t_i,r)=r$ at
the initial epoch $t=t_i$ from where the collapse commences. 
The time $t=t_s(r)$ corresponds to the formation of
the shell-focusing 
singularity at $R=0$, where all the matter shells collapse to a 
vanishing physical
radius.

Now let us assume that the mass function can be 
explicitly written as a function of the physical radius $R$ 
of the cloud and $t$, 
\begin{equation}
F(R,t)=R^3M(R)Q(t)
\label{eq:F}
\end{equation}
That is, we consider the class of mass functions $F$ 
to be separable in $R$ and $t$.
Apart from that it is general in the sense that $M$ is any $C^2$ 
function, whereas the function $Q(t)$ yet to be determined by the field 
equations is a suitably differentiable function of
$t$ for $t<t_{s_0}$, where $t=t_{s_0}$ is the time for
the occurrence of the central singularity.
One may consider equation (13) to be a somewhat strong assumption
on the nature of the mass function. However, our basic purpose
here is to construct a class of dynamical collapse models,
where the two main constraints of the homogeneous dust collapse mentioned 
above are relaxed, namely the density need not be homogeneous at 
the initial epoch and later as well during the collapse evolution,
and secondly we want to allow for non-zero pressures while
considering a dynamical collapse situation. As both these
purposes are met by the above form of mass function as we have
chosen here, it is adequate for our present purposes, as it allows 
us to construct explicit collapse models which are more general,
and which necessarily end up in a black hole as we shall see.

Another requirement that is frequently imposed
on physical grounds on the initial data from which the 
collapse evolves is that the 
physical variables such as the
density and pressures are taken to be smooth or analytic
functions at the initial surface.
We can then write,
\begin{equation}
M(R)=\frac{1}{3}+\frac{1}{5}M_2R^2+\cdots
\label{eq:M}
\end{equation}
Then from equation (\ref{eq:ein6}) we get
$\rho(r,t)=\rho(R,t)$ with,
\begin{equation}
\rho(R,t)=(3M+RM_{,R})Q(t)=A(R)Q(t)
\label{eq:rho}
\end{equation}
where the function $A(R)$ is given by,
\begin{equation}
A(R)=1+M_2R^2+\cdots
\label{eq:A}
\end{equation}
As seen from above, at the initial epoch $t=t_i$, 
the density function is given by,
\begin{equation}
\rho(r,t_i)=\rho_0(r)=Q(t_i)\left[1+M_2r^2+\cdots\right]
\label{eq:rhoini}
\end{equation}
Thus we see that the gradients of the density and pressures
of the cloud vanish at the center at the initial epoch
as required by the smoothness.
Also, for the density to diverge at the singularity, we must have,
\begin{equation}
\lim_{t\rightarrow t_{s_0}} Q(t)\rightarrow \infty
\label{eq:Q1}
\end{equation}
It follows that for the given mass function the perfect fluid condition
can be written as 
\begin{equation}
(k+1)QA+\frac{R}{\dot{R}}M\dot{Q}=0
\label{eq:pf1}
\end{equation}
The solution of the above equation determines the mass function
completely.

In order now to construct a class of collapsing solutions,
let us consider 
the case when $\nu=\nu(R)$, i.e. let the metric function 
$\nu$ be a function of the physical radius $R$ only. Again,
this restriction is good enough for us as it allows us to construct
the collapse models which include inhomogeneity and non-zero 
pressures, and which end up in black holes, generalizing the collapsing
Friedmann models. A further useful feature of these choices may
be considered to be that it allows to be incorporated 
a perfect fluid with a reasonable equation of state, rather
than any arbitrary
forms of pressures (e.g. a purely tangential pressure, while 
assuming that the radial pressures identically vanish) 
as is done some times when considering gravitational collapse.

One can now integrate equation 
(\ref{eq:ein7}) to get
\begin{equation}
\nu(R)=-\frac{k}{k+1}\ln[C_1A(R)]
\label{eq:nu}
\end{equation}
Here $C_1$ is a constant of integration. Now putting the value 
of $H(t,r)$ in equation (\ref{eq:ein3}) and simplifying we get,
\begin{equation}
R'\dot{G}-2\dot{R}\nu'G=0
\label{eq:G1}
\end{equation}
It is now possible to solve the above equation 
and the function $G$ has the form,
\begin{equation}
G(R)= A(R)^{-\frac{2k}{k+1}}
\label{eq:G2}
\end{equation}
In other words, the above forms of $\rho$, $\nu$ 
and $G$ solve the Einstein's equations (\ref{eq:ein2}), (\ref{eq:ein3}).
Obtaining now the function $Q(t)$ will complete the solution.
Putting in these functions in equation (\ref{eq:ein4}) we get,
\begin{equation}
\dot{R}=-C_2A^{\frac{-k}{k+1}}\sqrt{A^{\frac{-2k}{k+1}}-1+R^2MQ}
\label{eq:rdot1}
\end{equation}
where $C_2=C_1^{\frac{-k}{k+1}}$ is another constant. The negative
sign denotes the collapse condition $\dot{R}<0$.
Finally, substituting the values of the functions $A$ and $M$ we get,
ignoring the higher order terms, as we are interested to find a 
solution close to the singularity,
\begin{equation}
\dot{R}= - C_2R\left[1-\frac{2kR^2}{k+1}\right]\sqrt{\left[\frac{1}{3}+
\frac{1}{5}M_2R^2\right]Q-\frac{2kM_2}{k+1}} 
\label{eq:rdot2}
\end{equation}

We note that in the equation (24), the velocity $\dot R$ changes
sign at the value $R= \sqrt{(1+k)/2k}$. This would correspond to a 
class of dynamic perfect fluid models where there is a bounce at the
above value of the physical radius. However, since we are interested in 
the collapse models only presently, we do not consider this bouncing
branch of the solutions. This can be achieved by fixing the boundary
conditions suitably. For example, for the extreme value $k=1$, this
change corresponds to $R=1$. Now as we have the scaling $R=r$ at the 
initial epoch, this means that the boundary of the object $r=r_b$ 
at the initial epoch is to be given by $0<r_b<1$. Then at all later 
epochs this condition will be of course respected because $\dot R<0$, and 
the physical radius $R$ monotonically decreases with $t$ and will be
less than one for all shells at all future times. For smaller 
values of $k$, that is $k<1$, we of course have larger values of 
the boundary of the cloud available, and in the extreme case $k=0$,
i.e. the dust collapse models, we can have arbitrarily large $r_b$
without the velocity ever changing sign, or
the cloud can be as big as we want, and there will be no bounce at
all possible in the dust case.

Now let us solve the equation (\ref{eq:pf1}) 
close to the spacetime singularity at $R=0$.
In this approximation, 
$A(R)\rightarrow1$ and $M(R)\rightarrow\frac{1}{3}$.
Using these approximations and equation (\ref{eq:rdot2}) 
in equation (\ref{eq:pf1}) we get 
\begin{equation}
(k+1)Q-\frac{1}{3C_2\sqrt{\frac{Q}{3}-\frac{2k}{k+1}M_2}}\dot{Q}=0
\label{eq:q1}
\end{equation}
Considering that $M_2<0$ and solving the above equation with the 
boundary condition $Q(t_{s_0})\rightarrow\infty$ as pointed out
earlier, we get
\begin{equation}
Q(t)=-\alpha + \left[\alpha+\frac{2\alpha}{\left[\exp\{-\sqrt{3}C_2(k+1)
\alpha(t_{s_0}-t)\}-1\right]}\right]^2
\label{eq:q2}
\end{equation}
where,
\begin{equation}
\alpha=\sqrt{\frac{6k|M_2|}{k+1}}
\label{eq:alpha}
\end{equation}

Thus we see that the above $Q(t)$ is solution to Einstein's 
equations in the vicinity of the singularity with respect to the given 
forms of $\rho$, $\nu$ and $G$. Now we can also solve for the 
metric function $R$, which is the physical radius for the 
cloud. Using equation (\ref{eq:rdot2})
we get,
\begin{equation}
R(r,t)=f(r)e^{-B(t)}
\label{eq:R}
\end{equation}
where $f(r)$ is an arbitrary function of $r$. To avoid any shell
crossing singularity we consider $f$ to be an increasing function of
$r$ and since the area radius of the geometrical center of the 
cloud vanishes we must have $f(0)=0$. The function $B(t)$ is given as,
\begin{equation}
B(t)=C_2\int\sqrt{\frac{Q}{3}-\frac{2k}{k+1}M_2}dt
\label{eq:B}
\end{equation}

As noted earlier, the spacetime singularity occurs at $R=0$. 
We now need to decide if the singularity in the present case is
necessarily covered within an event horizon of gravity (which 
is the case of a black hole formation), or
it could be visible to faraway observers in the spacetime. The way to
decide this is to examine if there are any future directed families of 
null geodesics which go out to external observers in future, and which in
the past terminate at the singularity. If such families do exist, then 
the singularity is naked, which in principle can communicate with
outside world, and in the case otherwise we have a black
hole forming as the end state of collapse. We thus need to consider
the existence or otherwise of such families of paths from the
singularity.

With the form of $R$ as given above in equation (28), and 
as $\dot R<0$, the singularity  happens at a time 
$t=t_s$ when the physical radius for all the shells with different values 
of the comoving coordinate $r$ becomes zero. In other words,
there is a {\it simultaneous} collapse of all the shells to
singularity, and as $t\rightarrow t_{s_0}$ all the shells labeled by 
the coordinate $r$ collapse simultaneously to the singularity at $R=0$.
This necessarily gives rise to a covered central singularity at $R=0,r=0$,
as there are no outgoing future directed non-spacelike geodesics
coming out from the same. Because, if there were any such
outgoing geodesics, given by say $t=t(r)$ in the $(t,r)$ plane, which 
came out from $t=t_s,r=0$, then the time coordinate must increase 
along these paths, which is 
impossible as there is complete collapse at $t=t_s$, and there is no 
spacetime beyond that. Hence no values $t>t_s$ are allowed
within the spacetime which does not extend beyond the singularity.
Thus, the collapse gives rise necessarily to a black hole in the 
spacetime.

%\begin{equation}
%\int\frac{dt}{(t_{s_0}-t)^{C_2\sqrt{\beta}}}=f(r)
%\label{eq:null3}
%\end{equation}
%Now for the central singularity {\it i.e} in the limit $t\rightarrow t_{s_0}$,
%$r\rightarrow 0$, the {\it LHS} of the above equation diverges whereas 
%{\it RHS} tends to zero. Hence it follows that the assumption that
%there are outgoing null geodesics coming out from the central singularity
%leads to a contradiction. 
%Thus there can be no radial null geodesics coming out from the central 
%singularity, which is necessarily covered.    

Our main purpose here has been to generalize the homogeneous
dust collapse scenario to include non-zero pressures, and the 
inhomogeneities of density and pressures, which is physically more
realistic situation, while ensuring that the collapse end state 
is a black hole only. 
While the perfect fluid collapse models we considered here 
allow for inhomogeneities in density and pressure profiles, and do 
necessarily give rise to black holes as we have shown here, 
it should be 
kept in mind, as we have noted above, that the classes of mass 
functions and the velocity profiles for the collapsing shells as 
determined by the choice of metric function $\nu(R)$ considered 
here are rather special. It is an open problem to explore if we could
generalize these assumptions further, and if so to what extent, and still
continue to get black holes only and not the naked singularities as
the final end product of gravitational collapse. The point is 
it is known, for example, for inhomogeneous dust collapse 
\cite{dust} 
that as long as the inhomogeneities are within certain
limits, the result of collapse is a black hole. However, beyond that
criticality of inhomogeneities, the collapse could end in a naked
singularity. Investigating further specific, but physically 
more realistic models, may 
illustrate better such features of gravitationally 
collapsing configurations.

\end{document}